\documentclass[aps,prb,twocolumn,showpacs,preprintnumbers]{revtex4-1}

\usepackage[T1]{fontenc}
\usepackage[latin9]{inputenc}
\usepackage{amsmath}
\usepackage{graphicx}
\usepackage{amssymb}
\usepackage{braket}
\usepackage{textcomp}

\newcommand{\noun}[1]{\textsc{#1}}

\makeatother

\begin{document}
\title{Controlling the Dynamics of Quantum Mechanical Systems Sustaining Dipole-Forbidden Transitions via Optical Nanoantennas}

\author{Robert~Filter}
\author{Stefan~M\"{u}hlig}
\author{Toni~Eichelkraut}
\author{Carsten~Rockstuhl}
\author{Falk~Lederer}
\affiliation{Institute of Condensed Matter Theory and Solid State
Optics, Abbe Center of Photonics, Friedrich-Schiller-Universit\"{a}t Jena, Max-Wien-Platz 1, D-07743 Jena, Germany}

\begin{abstract}
We suggest to excite dipole-forbidden transitions in quantum-mechanical systems by using appropriately designed optical nanoantennas. The antennas are tailored such that their near-field contains sufficiently strong contributions of higher-order multipole moments. The strengths of these moments exceed their free space analogs by several orders of magnitude. The impact of such excitation enhancement is exemplarily investigated by studying the dynamics of a three-level system. It decays upon excitation by an electric quadrupole transition via two electric dipole transitions. Since one dipole transition is assumed to be radiative, the enhancement of this emission serves as a figure of merit. Such self-consistent treatment of excitation, emission, and internal dynamics as developed in this contribution is the key to predict any observable quantity. The suggested scheme may represent a blueprint for future experiments and will find many obvious spectroscopic and sensing applications.
\end{abstract}


\maketitle

\section{Introduction}
Optical nanoantennas have changed our perception of how light can interact with matter. If such a nanoantenna is made from a noble metal, it supports localized surface plasmon polaritons (LSPP) at distinct frequencies in the visible and near-infrared spectral range. LSPPs are excited when the electromagnetic radiation is resonantly coupled to the charge density oscillation in the metal. LSPPs allow the focusing of light into volumes (hot spots) inaccessible by classical optical devices. Moreover, the hot spot intensities may exceed the intensity of the external illumination by orders of magnitude.
These remarkable features render optical nanoantennas as prime candidates for controlling and improving the interaction of far-field light with other nanoscopic building blocks such as quantum dots, atoms, or molecules \cite{Frantsesson2001,Blanco2004b,Zuloaga10,Esteban2010,Sandoghdar2010,Giessen2011,Lee2011,Abb2011}. Here, we shall treat such nanoscopic building blocks at a quantum-mechanical level to fully grasp their internal dynamics and to appropriately describe their properties.

Many applications have been developed that exploit light interaction with such hybrid systems consisting of a nanoantenna and a nanoscopic building block. Pivotal examples can be found in the field of biosensing \cite{Zayats_NaMat_2009} and photovoltaics \cite{Atwater_NaLett_2008,Rockstuhl_APL_2009}.
In studies that considered individual hybrid systems, the most important finding consisted in showing that optical nanoantennas can modify the radiative decay rate of emitters \cite{Muskens2007}. By taking advantage of this effect, one can either enhance the emission rate or facilitate non-radiative decays. The interaction of light with such hybrid systems was first studied for highly symmetric optical nanoantennas as, e.g., metallic nanospheres. However, more complex antennas can be equally considered \cite{Filter2011circular}. In such cases, the emission characteristics of these hybrid systems might be entirely governed by the optical nanoanantenna and may strongly deviate from the radiation pattern of emitters in free space \cite{vanHulst2010}. These new engineering possibilities pave the way for the development of highly directed single photon sources and other applications.\cite{vanHulst2008,Esteban2010}

\begin{figure}
\begin{centering}
\includegraphics[width=80mm]{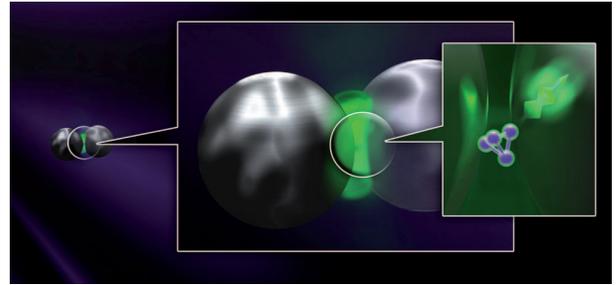}
\par\end{centering}
\caption{\label{fig:ToC}
Left: a plane wave (purple) scatters two close silver nanospheres. Middle: molecules in-between undergo a quadrupolar excitation resulting in a luminescence signal at a lower wavelength (green). Right: excitation and emission of a single molecule.}
\end{figure}

However, in all the aforementioned studies the interaction of optical nanoantennas with the quantum-mechanical system has been discussed by resorting to more or less appropriate approximations. To be specific, apart from a few exceptions\cite{Zurita-Sanchez2002,Zia2011,Martin2012}, the effect of higher order multipole fields in the vicinity of the antenna has been restricted to a pure electric dipole field; although experimental indications for the influence of higher order multipole fields have been reported in the literature.\cite{Moskovits1982,Kawazoe2002}
The restriction to electric dipole fields seemed to be an obvious and quite reasonable approximation, since in free space the interaction of a quantum-mechanical system with higher-order multipole fields is orders of magnitude weaker than compared to that of an electric dipole field; as discussed in detail below. Moreover, none of the aforementioned studies considers the modification of the dynamics of the quantum-mechanical system subject to a higher-order excitation. However, subsequent calculations will show that this is of crucial importance.

With the self-imposed limitation to consider the field close to an optical nanoantenna as to be electric dipolar, the potential of optical nanoantennas is unnecessarily restricted because there is a much larger flexibility in tailoring their near-field. In this work the potential of optical nanoantennas to significantly enhance higher-order multipole fields in the vicinity of the nanoantenna is considered. The enhancement is such strong that it can be essential in the interaction with a quantum-mechanical system.
Most notably, it enables the significant excitation of transitions which are typically considered \textit{forbidden} in free-space.

To analyse such processes, a theoretical framework necessary to calculate non-dipolar transitions of quantum-mechanical multilevel systems in the vicinity of optical nanoantennas is established in this contribution. These non-dipolar transitions, often termed as \textit{forbidden} ones, can be tremendously enhanced due to large higher-order field components \cite{Zurita-Sanchez2002,Tojo2005,Deguchi2009,Martin2012}. In passing we note that these higher-order components may also be used to excite dark modes in plasmonic systems using strong field gradients.\cite{Stockman2001} Such modes may also couple to quantum systems \cite{Li2005,Yan2008,Liu2009}.
The emission in the vicinity of nanoantennas is also strongly modified with respect to free space \cite{Welsch2006}. Thus, to understand the dynamics of a quantum system it is essential to also account for these modified emission characteristics. As an example, a three-level system that is excited through an electric quadrupole transition and relaxes via two consecutive electric dipole transitions is investigated in detail in this contribution.

The paper is organized as follows.
At first, a mathematical framework to discuss the enhancement of higher-order multipole fields in the vicinity of optical nanoantennas is introduced. For the sake of definiteness, the discussion is exemplified at a suitably chosen nanoantenna, i.e. a nanoantenna which strongly enhances the quadrupole field.
Next, the modification of excitation rates in quantum mechanical systems due to this quadrupolar enhancement will be investigated. Finally, the hybrid system consisting of an optical nanoantenna and a quantum-mechanical three-level system will be studied in detail. A detailed appendix provides further explanations and explicit derivations of formulas used in the main body of the manuscript.

It is the purpose of this contribution to show that a properly designed nanoantenna can excite dipole \textit{forbidden} transitions in three-level systems due to the enhanced higher-order multipole fields. It will be shown that the dynamics of the system are strongly altered by the presence of the nanoantenna and cannot be understood by the quadrupolar enhancement alone.
Although, only shown here at the example of a three-level system, more complicated quantum systems can equally be considered and the framework presented herein can be adjusted to different possible experimental configurations.

\section{Enhancement of Higher-Order Multipole Fields}
Previously, higher-order multipole transitions in hydrogen-like atoms have usually been assumed to be \textit{forbidden} (except for extreme situations \cite{Bhattacharya2003}), since the corresponding contributions as provided by an excitation field usually used in most cases, i.e. a plane wave, are much weaker than the dipolar ones.
This can be estimated from a back of the envelope calculation that assumes some characteristic values. A stricter derivation can be found in, e.g., Ref.~\onlinecite{Hertel2007}.
The electric field of a plane wave varies spatially as $\exp\left[\mathrm{i}(\mathbf{k}\cdot \mathbf{x})\right]\approx 1 + \mathrm{i}(\mathbf{k}\cdot \mathbf{x})$ in the limit of $\mathbf{k}\cdot \mathbf{x}\rightarrow 0$. Here $\mathbf{k}$ is the wavevector that can be assumed as  $\left|\mathbf{k}\right|\approx10^7\,\mathrm{m}^{-1}$ for visible light and the characteristic spatial extent of the atomic system being $\left<\left|\mathbf{x}\right|\right>\approx Z a_0\approx 10^{-10}$~m for hydrogen-like atoms. Only the first term in the expansion has been retained, corresponding to the electric dipole component. In most cases this approximation is reasonable since the first order term in the Taylor expansion is three orders of magnitude larger than the second term, which is attributed to both the electric quadrupole and magnetic dipole fields. Hence, for the given spectral domain and the usually considered spatial extent of the atomic system, the excitation rates induced by the local quadrupole field of a plane wave are orders of magnitude smaller than those transitions induced by the electric dipole field. For this reason, quadrupole transitions are usually said to be inaccessible, i.e. they are forbidden. Clearly, components of octupolar or higher order are even weaker.

In the presence of an optical nanoantenna the situation changes dramatically. Such plasmonic structures support highly localized near-fields that are characterized by huge gradients. Using a multipole expansion in spherical coordinates, the electric field in a coordinate system with origin $\mathbf{r}_0$ can be expressed as
\begin{eqnarray}
\mathbf{E}\left(\mathbf{x},\omega\right) &=& \sum_{m,n} [ p_{mn}\left(\omega;\mathbf{r}_0\right)\mathbf{N}_{mn}(\mathbf{x}-\mathbf{r}_0,\omega) + \nonumber \\
& &\quad \ \, q_{mn}\left(\omega;\mathbf{r}_0\right)\mathbf{M}_{mn}(\mathbf{x}-\mathbf{r}_0,\omega)]
\label{eq:Multipole_Field_Representation}
\end{eqnarray}
with vector spherical harmonics $\mathbf{N}_{mn}$ and $\mathbf{M}_{mn}$ following the notation of Ref.~\onlinecite{Xu1995}. Equation~\ref{eq:Multipole_Field_Representation} is identical to a multipole expansion in spherical coordinates (except some prefactors) with $p_{mn}$ and $q_{mn}$ being the complex electric and magnetic multipole coefficients, respectively \cite{Rockstuhl_PRB_2011}.
The order $n=1$ corresponds to electric and magnetic dipoles, whereas $n=2$ corresponds to quadrupoles and so on.

For illustration purposes, in what follows the focus shall lie on the electric quadrupole. Moreover, to simplify the subsequent discussion, the more familiar Cartesian quadrupole components $Q_{ij}$ will be used. They can be obtained from the electric quadrupole coefficients $p_{m2}$ using linear transformations \cite{Muehlig2011}.

\begin{figure}
\begin{centering}
\includegraphics[width=80mm]{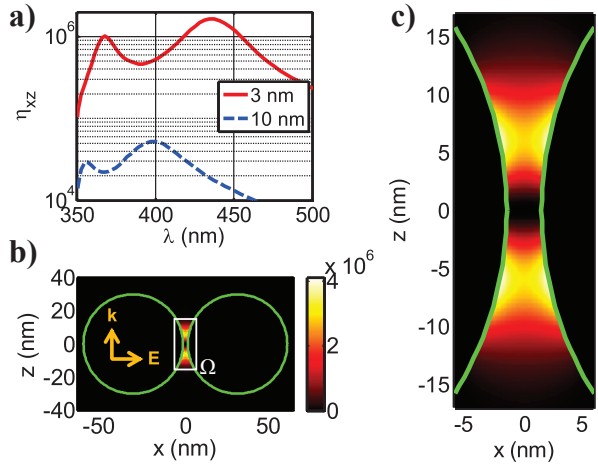}
\par\end{centering}
\caption{\label{fig:local_Q}
(Color online)
a) Integrated enhancement factor $\eta_{xz}$ as a function of excitation wavelength for two separations of the dimer nanospheres under plane wave illumination as sketched in b). The area $\Omega$ pertinent for integration was chosen to be $\Delta z=30$~nm and $\Delta x = 13$~nm and fixed at $y=0$ [white box in b)]. A broadband enhancement of about six orders of magnitude for a separation of $d=3$~nm can be observed which peaks at $370$~nm and $437$~nm. b) The local quadrupole enhancement $\eta_{xz}^{\mathrm{loc}}(\mathbf{r}_0)$ in the $x$-$z$-plane for $d=3$~nm and at $\lambda=437$~nm. c) Zoomed view of b) (color bar is maintained) where two symmetric stripes of maximum enhancement can be observed.}
\end{figure}

As a referential optical nanoantenna that supports strong local electric quadrupole fields, an optical nanoantenna consisting of two strongly coupled silver nanospheres, sometimes termed as dimer, is investigated.
The two silver nanospheres have a radius of $30$~nm and are separated by either $3$~nm or $10$~nm. With such a separation, quantum effects and/or a possible nonlocal material properties of the silver nanosphere may only constitute a minor contribution which we do not consider here \cite{GarciadeAbajo2008,McMahon2009}; material parameters were taken from Ref.~\onlinecite{Johnson1972}.
The assumed geometry is in reach of state-of-the-art fabrication techniques and can even be scaled to large arrays of strongly coupled nanospheres.\cite{Cunningham2011}
It is well established that the scattering cross section of such a dimer structure exhibits a strong quadrupole contribution for a plane wave illumination direction parallel to the connecting line of the nanospheres \cite{Muehlig2011}.
However, the enhancement of local quadrupole fields, as probed by the quantum mechanical system, was found to be strongest if the two nanospheres are illuminated perpendicular to the connecting line (chosen to be the $x$-axis) with a polarization of the electric field parallel to it (see Figure.~\ref{fig:local_Q}). However, the quadrupole contribution to the far field is negligible for this illumination scenario.

Considering plane wave illumination in the given coordinate system, the only nonvanishing component of the related quadrupole tensor in free space is $Q_{xz}^{\mathrm{fs}}$. This particular coefficient is linked to the multipole expansion coefficient as used in Eq.~\ref{eq:Multipole_Field_Representation} via  $Q_{xz}=\frac{i}{\sqrt{6}}(p_{-12}- p_{12})$. Therefore, the local quadratic enhancement of a quadrupole field may be defined as the ratio $\eta_{ij}^{\mathrm{loc}}(\mathbf{r}_0) = \left|Q_{ij}^{\mathrm{na}} (\mathbf{r}_0)/Q_{xz}^{\mathrm{fs}} (\mathbf{r}_0)\right|^2\,$. One may also define the integrated enhancement factor $\eta_{ij} =\int_\Omega \eta_{ij}^{\mathrm{loc}} \left(\mathbf{r}_0\right)dV$ with respect to a certain domain $\Omega$. Note that for enhancements obeying certain symmetries it is convenient to regard a lower dimensional integration. Throughout this manuscript, the superscripts '$\mathrm{fs}$' and '$\mathrm{na}$' designate the free-space and the nanoantenna scenario, respectively.

Figure~\ref{fig:local_Q}~a) shows $\eta_{xz}$ as a function of excitation wavelength for two different separations of the nanospheres. The dimer is illuminated by a plane wave as sketched in Fig.~\ref{fig:local_Q}~b). It can clearly be seen that for a separation of $3$~nm $\eta_{xz}$ has a maximum of $1.6\times10^6$ at $437$~nm. Because it is strongest, we will exclusively focus on this resonance. For comparison Fig.~\ref{fig:local_Q}~a) also shows $\eta_{xz}$ for a nanosphere separation of $10$~nm. The integrated enhancement factor is approximately two orders of magnitude less than compared to the $3$~nm separation. Therefore, it is obvious that the enhancement of the electric quadrupole contribution with respect to the near field stems from the strong coupling of the two nanospheres, which critically depends on their separation. All other components of $\eta_{ij}$ are orders of magnitude less than $\eta_{xz}$ and shall therefore be neglected here. Figure~\ref{fig:local_Q}~b) shows $\eta_{xz}^{\mathrm{loc}}\left(\mathbf{r}_0\right)$ in the $xz$ plane for a  $3$ nm separation at a wavelength of $437$ nm. The enhancement of the electric quadrupole contribution takes place in a narrow spatial domain between the surfaces of the two nanospheres. In Fig.~\ref{fig:local_Q}~c), which displays a zoomed view of b), it can be observed, that $\eta_{xz}^{\mathrm{loc}}$  consists of two symmetric stripes of approximately $5$~nm width.

\section{Forbidden Transitions}

After having considered the enhancement of the local electric quadrupole field near the nanoantenna, the resulting enhancement of quadrupole transition rates will now be examined.
The transition rates from the $i^\mathrm{th}$ to the $j^\mathrm{th}$ eigenstate of a non-relativistic quantum-mechanical system with unperturbed \noun{Hamiltonian} $H_0$ follows \noun{Fermi}'s golden rule
\begin{equation}
\Gamma_{ij} = \frac{2\pi}{\hbar^2} \left| \left\langle i\right| V \left|j\right\rangle \right|^2
\left\{ \delta\left(\omega_{ij}-\omega \right)+ \delta\left(\omega_{ij}+\omega \right) \right\}\ .
\label{eq:Fermi}
\end{equation}
Here, the time-harmonic interaction potential $V_I = e^{-\mathrm{i}\omega t}V(\mathbf{x}) + e^{\mathrm{i}\omega t}V^\dagger(\mathbf{x})$ is assumed to be a small perturbation. Within this framework, the action of an electromagnetic field on an electron with momentum operator $\mathbf{p}$ is given by the minimal coupling interaction potential $V_I =-\frac{e}{m}\mathbf{A}\cdot\mathbf{p} + \text{h.c.}$ \cite{Hertel2007}, see also Appendix~\ref{sub:Minimal_Coupling_Interaction_Potential}. Here, the weak-field approximation is applied and hence the ponderomotive potential $\frac{e^2}{2m}\mathbf{A}^2$ shall be neglected. The electromagnetic fields are related to the vector potential $\mathbf{A}$ by $\mathbf{B} = \mathrm{curl}\mathbf{A}$ and $\mathbf{E} = \mathrm{i}\omega\mathbf{A}-\mathrm{grad}U$.
Choosing the \noun{Coulomb} gauge where $U \equiv 0$ can always be achieved, the electric field is given by $\mathbf{E} = \mathrm{i}\omega\mathbf{A}$.

Furthermore, a decomposition of the interaction potential yields
$V_I = -\frac{\mathrm{i}e}{\hbar}\left\{ \left[H_{0},\mathbf{A}\cdot\mathbf{x}\right]+\left[\mathbf{A},H_{0}\right]\mathbf{x}\right\}$.
In this form, the terms can be directly interpreted as electric and magnetic coupling terms, $V = V_e + V_m$ with
\begin{eqnarray}
V_e =-\frac{e}{\hbar\omega}\left[H_{0},\mathbf{E}\cdot\mathbf{x}\right] &,&
V_m = \frac{\mu_{B}}{\hbar}\mathbf{L}\cdot\mathbf{B} \label{eq:el_magn_coupling}
\end{eqnarray}
where $\mu_{B}=\frac{e\hbar}{2m}$ is \noun{Bohr}'s magneton and $\mathbf{L}=\mathbf{x}\times\mathbf{p}$ denotes the angular momentum operator. A derivation of this decomposition can be found in Appendix~\ref{sub:Decomposition}.

Without loss of generality, the analysis shall be restricted to electric transitions.
Referring to Eq.~\ref{eq:Multipole_Field_Representation} and the relation $\mathbf{M}_{mn}(\mathbf{x})\cdot\mathbf{x}=0$, the electric interaction potential $V_e$ can be fully characterized by the complex multipole coefficients $p_{mn}$ and the related vector spherical harmonics $\mathbf{N}_{mn}$. Using the definition of the $\mathbf{N}_{mn}$ following the notation of Ref.~\onlinecite{Xu1995}, one finds the explicit expression
\begin{equation}
\mathbf{E}\cdot\mathbf{x} = \sum_{m,n}p_{mn}\frac{n(n+1)}{k}
j_n(kr)P_n^m\left(\cos\theta\right)\exp(\mathrm{i}m\varphi)
\label{eq:E_times_x}
\end{equation}
where $j_n$ denotes spherical \noun{Bessel} functions and $P_n^m$ associated \noun{Legendre} polynomials. The latter expression holds in the coordinate system of the quantum system, i.e. for $\mathbf{r}_0=0$.

As mentioned earlier, this spherical representation can also be transformed to a \noun{Cartesian} one and the moments can locally be related to the coefficients of a \noun{Taylor} expansion, $\mathbf{E}\cdot\mathbf{x}\propto \sum_{n}\sum_{i\dots k}^n Q_{i\dots k}x^i\dots x^k$. Then, it becomes evident why the quadrupole coupling term ($n=2$) is related to the near field gradients of the electric field: $Q_{ij}\propto\partial_i\partial_j \left( \mathbf{E}\cdot\mathbf{x}\right)$. However, care has to be taken regarding the normalization of the $Q_{ij}$ which is not consistent throughout the literature. For the calculations in the main body of the manuscript, we have chosen the notation as given in Ref.~\onlinecite{Muehlig2011}.

Combining Eqs. \ref{eq:Multipole_Field_Representation} - \ref{eq:el_magn_coupling}, the transition rates, related to different multipole orders, can be computed directly from a given field distribution. Hence, the transition rate at the atomic site $\mathbf{r}_0$, resulting from the electric potential $V_e$, is proportional to the square modulus of a linear combination of the multipole coefficients $p_{mn}$, i.e.
\begin{equation}
\Gamma_{ij}(\mathbf{r}_0) = \frac{2\pi e^2}{\hbar^2}\left|\sum_{n,m}p_{mn}(\mathbf{r}_0) \left\langle i\right| \mathbf{N}_{mn}\cdot\mathbf{x} \left|j\right\rangle
\right|^2\delta\left(\omega_{ij}\pm\omega\right) \ .
\label{eq:KeyMessage}
\end{equation}

Equation~\ref{eq:KeyMessage} establishes the link between the enhancement of higher order electric transitions in a quantum-mechanical system and contributions of higher order multipoles to the local field as investigated in Fig.~\ref{fig:local_Q}. Therefore, the previously investigated optical nanoantenna can be used to increase the strength of electrical quadrupole transitions in a quantum mechanical system.

\section{Modified Emission Characteristics}

After having demonstrated that near optical nanoantennas the quadrupole field can be enhanced by many orders of magnitude, the focus will now be directed to the effect of these multipole contributions on the dynamics of the entire quantum-mechanical system. For simplicity, a three-level system will be considered in the following, which is excited through an electric quadrupole transition and decays via two consecutive electric dipole transitions, see Fig.~\ref{fig:Gamma_compare} a).
It must be noted that it is reasonably assumed herein that only the quadrupolar fields contribute to the quadrupole transition, i.e. the transition is dipole-forbidden. If this assumption does not hold, the interaction to other enhanced multipole components of the field has to be taken into account following Eq.~\ref{eq:KeyMessage}. As an example, a comparison to local dipolar enhancements is given in Appendix~\ref{sub:Dipole_Enhancement}.
In the subsequent discussion it will be shown how the emission characteristics of the entire system are modified in the presence of the nanoantenna. As a system of reference, a dimer with a distance of $3$~nm between the silver nanospheres has been chosen.

\subsection{Dynamics of the Three-Level System}

The dynamics of the three-level system are governed by the following rate equations. They describe the population of the three energy levels, i.e.
\begin{eqnarray*}
\dot{n}_{0} & = & \gamma_{10}\cdot n_{1}-\Gamma_{02}\cdot n_{0} \ \mathrm{,} \\
\dot{n}_{1} & = & \gamma_{21}\cdot n_{2}-\gamma_{10}\cdot n_{1} \ \mathrm{,\ and} \\
\dot{n}_{2} & = & \Gamma_{02}\cdot n_{0}-\gamma_{21}\cdot n_{2} \ .
\end{eqnarray*}
Here, the $\gamma_{ij}$ denote the spontaneous decay rates from the $i^\mathrm{th}$ to the $j^\mathrm{th}$ level, whereas $\Gamma_{02}$ denotes the excitation rate of the quadrupole transition. The quadrupole transition takes place at a wavelength of $\lambda_{02}=437$~nm, for which the previously investigated optical nanoantenna provides the maximal enhancement of an electric quadrupole field [see Fig.~\ref{fig:local_Q}~a)].

In the system under investigation, quadrupolar emission processes at $\lambda_{02}$ are neglected \cite{Klimov1996PRA}, because the spontaneous decay from the second to the first level is assumed to be much faster. The emission wavelength of the first dipole transition is assumed to be $\lambda_{21} = 3.47$~$\mathrm{\mu}$m; consequently, the second dipole transition takes place at $\lambda_{10} = 500$~nm. For both dipole transitions, stimulated processes are neglected.

In free space and under plane wave illumination with unit intensity, the ratios $\Gamma_{02}^{\mathrm{fs}}/\gamma_{21}^{\mathrm{fs}} = 10^{-5}$ and $\gamma_{10}^{\mathrm{fs}}/\gamma_{21}^{\mathrm{fs}} = 10^{-2}$ were chosen.
These rates correspond to real physical systems as outlined in Appendix~\ref{sec:choices}.

To investigate the enhancement of the efficiency of the quadrupole transition due to the nanoantenna, the amount of light spontaneously emitted at $\lambda_{10} = 500$~nm per unit time shall serve as a figure of merit. Therefore, the rate equations are solved in equilibrium.
In this regime, luminescence is given by $\dot{n}_{1}^{\mathrm{rad}} = \gamma_{10}^{\mathrm{rad}}\,n_1$, where $\gamma_{10}^{\mathrm{rad}}$ is the spontaneous radiative decay rate. The total decay rate is given by the relation $\gamma_{10}=\gamma_{10}^{\mathrm{rad}}+\gamma_{10}^{\mathrm{nonrad}}$, where $\gamma_{10}^{\mathrm{nonrad}}$ describes the nonradiative decay rate.

Note, in this study in free space $\gamma^{\mathrm{fs}}_{10}=\gamma_{10}^{\mathrm{rad},\mathrm{fs}}$, and $\gamma_{10}^{\mathrm{nonrad},\mathrm{fs}}=0$. On the other hand, in the presence of the nanoantenna $\gamma_{10}^{\mathrm{nonrad},\mathrm{na}}$ is merely determined by the absorption in the metal. From the rate equations in steady state and $1=n_0+n_1+n_2$, one can solve for $n_1$ finding
\begin{eqnarray}
\dot{n}_{1}^{\mathrm{rad}}=\gamma_{10}^{\mathrm{rad}}\frac{\gamma_{21}\,\Gamma_{02}}
{\gamma_{10}\,\gamma_{21}+\gamma_{21}\,\Gamma_{02}+\gamma_{10}\,\Gamma_{02}}\label{eq:n1_dot_rad}
\end{eqnarray}
which holds both in free space and in the vicinity of the nanoantenna.
Since the quadrupole enhancement of the electric field varies spatially, the same holds for $\Gamma_{02}$. Furthermore, the presence of the nanoantenna alters the local density of states. Thus, the spatial dependence of the spontaneous decay rates $\gamma_{ij}$ also has to be considered.

Within the weak atom-field coupling regime, the spontaneous decay rates are given as $\gamma_{ij}=\frac{2\omega^{2}}{\hbar\epsilon_{0}c^{2}}\, \left| \mathbf{d}_{ij}\right|^2\,\Im\left[G(\mathbf{r}_{0},\mathbf{r}_{0},\omega_{ij})\right]$, where $G$ is the component of the \noun{Green}'s tensor corresponding to the direction of the dipole moment $\mathbf{d}_{ij}$ of the specific transition~\cite{Welsch2006}. The \noun{Green}'s tensor is calculated at the position $\mathbf{r}_{0}$ of the quantum system.
Using this relation, the decay rates in the vicinity of the nanoantenna $\gamma_{ij}^{\mathrm{na}}$ are related to the decay rates in free space $\gamma_{ij}^{\mathrm{fs}}$ via $$\gamma_{ij}^{\mathrm{na}}/\gamma_{ij}^{\mathrm{fs}} = \Im\left[G^{\mathrm{na}}(\mathbf{r}_{0},\mathbf{r}_{0},\omega_{ij})\right]/
\Im\left[G^{\mathrm{fs}}(\mathbf{r}_{0},\mathbf{r}_{0},\omega_{ij})\right]\ .$$
Since $Q_{xz}$ is the dominant quadrupole component, the emission from the dipole transitions is assumed to be either $x$- or $z$-polarized; hence $\gamma_{ij}^{\mathrm{na}}$ is the arithmetic mean of these two contributions.
\begin{figure}
\begin{centering}
\includegraphics[width=80mm]{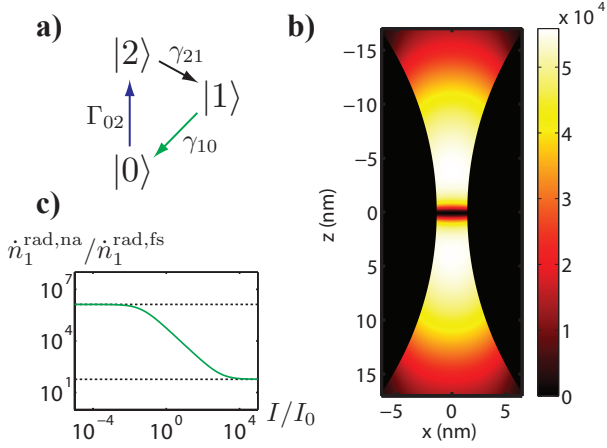}
\par\end{centering}
\caption{\label{fig:Gamma_compare}
a) Scheme of the quantum mechanical system that is placed in the vicinity of the nanoantenna.
b) Local enhancement of the luminescence, i.e. $\dot{n}_{1}^{\mathrm{rad},\mathrm{na}}/\dot{n}_{1}^{\mathrm{rad},\mathrm{fs}}$, in the $x$-$z$-plane. Because of saturation, the enhancement is not as strong as expected from the quadrupole enhancement alone.
c) Luminescence enhancement as a function of the intensity at the location $x=0$~nm, $z=4$~nm.
The dashed lines correspond to predictions for low and high intensities as discussed in section~\ref{sub:Excitation_Luminescence_Limiting} (i) and (ii).}
\end{figure}

In the subsequent discussion, $\dot{n}_{1}^{\mathrm{rad}}$ is chosen as the figure of merit, assessing how efficiently the emission at $\lambda_{10}$ can be raised due to the enhancement of the electric quadrupole fields.
Figure~\ref{fig:Gamma_compare}~b) shows $\dot{n}_{1}^{\mathrm{rad},\mathrm{na}}/\dot{n}_{1}^{\mathrm{rad},\mathrm{fs}}$. It can be seen that $\dot{n}_{1}^{\mathrm{rad},\mathrm{na}}$ can be enhanced by over four orders of magnitude relative to free space. As expected, in regions where the quadrupole enhancement is the strongest, the luminescence is strongly enhanced as well. However, due to saturation effects and nonradiative losses, characteristic values of the luminescence rates are about two orders of magnitude lower than the quadrupole enhancement. This can be understood in terms of different limiting cases of Eq.~\ref{eq:n1_dot_rad} and will be discussed in the following.

\subsection{The Effect of Excitation Intensities on the Luminescence Enhancement\label{sub:Excitation_Luminescence_Limiting}}

Naturally, the chosen ratios of the rates affect the luminescence enhancement.
However, the excitation rate $\Gamma_{02}$ depends not only on the geometry but also on the
intensity. Using Eq.~\ref{eq:n1_dot_rad}, one may estimate the luminescence enhancement
$\dot{n}_{1}^{\mathrm{rad,na}}/\dot{n}_{1}^{\mathrm{rad,fs}}$
by the main contributions given in the nominator for free space and
in the vicinity of the nanoantenna with respect to different intensities:
\begin{widetext}
\begin{eqnarray}
\frac{\dot{n}_{1}^{\mathrm{rad,na}}}{\dot{n}_{1}^{\mathrm{rad,fs}}} & = & \underbrace{\frac{\gamma_{10}^{\mathrm{rad,na}}}{\gamma_{10}^{\mathrm{rad,fs}}}
\cdot\frac{\Gamma_{02}^{\mathrm{na}}}{\Gamma_{02}^{\mathrm{fs}}}}_{\begin{array}{c}
\mathrm{Purcell\ effect\; and}\\
\mathrm{quadrupole\ enhancement}
\end{array}}\cdot\underbrace{\frac{\gamma_{10}^{\mathrm{fs}}\,\gamma_{21}+\gamma_{21}\,\Gamma_{02}^{\mathrm{fs}}+\gamma_{10}^{\mathrm{fs}}\,\Gamma_{02}^{\mathrm{fs}}}{\gamma_{10}^{\mathrm{na}}\,\gamma_{21}+\gamma_{21}\,\Gamma_{02}^{\mathrm{na}}+\gamma_{10}^{\mathrm{na}}\,\Gamma_{02}^{\mathrm{na}}}}_{\begin{array}{c}
\mathrm{dynamics\ of\ quantum\ system}\end{array}}\ .\label{eq:luminescence_enhancement_Aufteilung}
\end{eqnarray}
\end{widetext}
In this form it tends to be evident that not only the enhancement of the radiative rate of the luminescing transition, $\gamma_{10}^{\mathrm{rad,na}}/\gamma_{10}^{\mathrm{rad,fs}}$, has to be taken into account as well as the quadrupole enhancement $\Gamma_{02}^{\mathrm{na}}/\Gamma_{02}^{\mathrm{fs}}$ but also the dynamics of the quantum system given in the last term in Eq.~\ref{eq:luminescence_enhancement_Aufteilung}. In passing we note that the enhancement of radiative rates in an environment is usually termed Purcell effect.

As discussed before, the excitation rate in free space
at an intensity $I_{0}$ was assumed to be $\Gamma_{02}^{\mathrm{fs}}\left(I=I_{0}\right)=10^{-5}\gamma_{21}^{\mathrm{fs}}$.
Also $\gamma_{21}=10^{2}\gamma_{10}^{\mathrm{fs}}$ has been chosen
implying $\Gamma_{02}^{\mathrm{fs}}\left(I=I_{0}\right)=10^{-3}\gamma_{10}^{\mathrm{fs}}$.
Furthermore, in the vicinity of the nanoantenna $\gamma_{10}$ gets
also hugely enhanced and $\gamma_{10}^{\mathrm{na}}\gg\gamma_{21}$
holds. Equation~\ref{eq:luminescence_enhancement_Aufteilung} can be understood for different limiting cases of low, intermediate, or high intensities. This is done in the following and will explain the results found in Fig.~\ref{fig:Gamma_compare} in detail.

\textbf{(i) Low Intensities.} First of all, one might consider the case of very low intensity, $\Gamma_{02}\ll\gamma_{ij}$
both in free space and close to the nanoantenna. Then, one has
\begin{eqnarray*}
\frac{\dot{n}_{1}^{\mathrm{rad,na}}}{\dot{n}_{1}^{\mathrm{rad,fs}}} & \approx & \frac{\gamma_{10}^{\mathrm{rad,na}}}{\gamma_{10}^{\mathrm{rad,fs}}}\cdot\frac{\Gamma_{02}^{\mathrm{na}}}{\Gamma_{02}^{\mathrm{fs}}}\cdot\frac{\gamma_{10}^{\mathrm{fs}}\,\gamma_{21}}{\gamma_{10}^{\mathrm{na}}\,\gamma_{21}}
  =  \frac{\gamma_{10}^{\mathrm{rad,na}}}{\gamma_{10}^{\mathrm{na}}}\cdot\frac{\Gamma_{02}^{\mathrm{na}}}{\Gamma_{02}^{\mathrm{fs}}}\\
 & = & \mathrm{antenna\ efficiency}\ \times\ \mathrm{quadrupole\ enhancement}
\end{eqnarray*}
since in free space $\gamma_{10}^{\mathrm{fs}}=\gamma_{10}^{\mathrm{rad,fs}}$.
One can see that the antenna efficiency $\gamma_{10}^{\mathrm{rad,na}}/\gamma_{10}^{\mathrm{na}}$ is a limiting factor which will also naturally hold for the other limiting cases.
Furthermore, the luminescence enhancement is basically given by the quadrupole enhancement and can be calculated without consideration of the internal dynamics. Thus one can anticipate that the enhancement is most pronounced in this intensity regime. If the intensity is stronger, one can expect that the internal dynamics act as a bottleneck which will be examined in the following.

\textbf{(ii) High Intensities.} Now the intensity will be assumed to be so high that in free space and in the vicinity of the nanoantenna $\Gamma_{02}\gg\gamma_{ij}$ holds.
Then,
\begin{eqnarray*}
\frac{\dot{n}_{1}^{\mathrm{rad,na}}}{\dot{n}_{1}^{\mathrm{rad,fs}}} & \approx & \frac{\gamma_{10}^{\mathrm{rad,na}}}{\gamma_{10}^{\mathrm{rad,fs}}}\cdot\frac{\Gamma_{02}^{\mathrm{na}}}{\Gamma_{02}^{\mathrm{fs}}}\cdot\frac{\gamma_{21}\,\Gamma_{02}^{\mathrm{fs}}}{\gamma_{10}^{\mathrm{na}}\,\Gamma_{02}^{\mathrm{na}}}
  =  \frac{\gamma_{10}^{\mathrm{rad,na}}}{\gamma_{10}^{\mathrm{na}}}\cdot\frac{\gamma_{21}}{\gamma_{10}^{\mathrm{rad,fs}}}\\
 & = & \mathrm{antenna\ efficiency\ }\times10^{2}
\end{eqnarray*}
consistent with Fig.~\ref{fig:Gamma_compare}~c) for $I\gg I_{0}$.
Interestingly the enhancement is now independent of the quadrupole
enhancement. This results from the fact that in this intensity regime
also in free space the excitation is the fastest process.

\begin{figure}
\begin{centering}
\includegraphics[width=40mm]{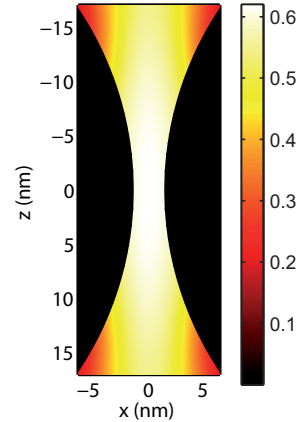}
\par\end{centering}
\caption{\label{fig:antenna_efficiency}
The antenna efficiency $\gamma_{10}^\mathrm{rad,na}/\gamma_{10}^\mathrm{na}$. It acts as a natural limiting factor for the luminescence enhancement.
Noteworthy, with respect to the actual form of the quadrupole enhancement outlined in Fig.~\ref{fig:local_Q}~c),
a perfect agreement of the luminescence enhancement predicted by equation Eq.~\ref{eq:intermediate_intensities} and the results given in Fig.~\ref{fig:Gamma_compare}~b) can be seen.
}
\end{figure}

\textbf{(iii) Intensities comparable to $I_0$.} Finally, one may consider an intermediate case where $I\approx I_{0}$.
Here, $\Gamma_{02}^{\mathrm{fs}}\ll\gamma_{21}$ but one may still assume that $\Gamma_{02}^{\mathrm{na}}\gg\gamma_{21}$. This is the case in the central
region of the nanoantenna which can be seen in Fig.~\ref{fig:local_Q}~c)
where the quadrupolar enhancement is shown to be in the order of $10^{6}$. Thus,
\begin{eqnarray}
\frac{\dot{n}_{1}^{\mathrm{rad,na}}}{\dot{n}_{1}^{\mathrm{rad,fs}}} & \approx & \frac{\gamma_{10}^{\mathrm{rad,na}}}{\gamma_{10}^{\mathrm{rad,fs}}}\cdot \frac{\Gamma_{02}^{\mathrm{na}}}{\Gamma_{02}^{\mathrm{fs}}}\cdot \frac{\gamma_{10}^{\mathrm{fs}}\,\gamma_{21}}{\gamma_{10}^{\mathrm{na}}\,\Gamma_{02}^{\mathrm{na}}}
  =  \frac{\gamma_{10}^{\mathrm{rad,na}}}{\gamma_{10}^{\mathrm{na}}}\cdot \frac{\gamma_{21}}{\Gamma_{02}^{\mathrm{fs}}}\nonumber\\
 & = & \mathrm{antenna\ efficiency\ }\times10^{5}\ .\label{eq:intermediate_intensities}
\end{eqnarray}
It is important to note that the results outlined in Fig.~\ref{fig:Gamma_compare}~b) correspond to the discussed intermediate intensity limit. There, maximum luminescence enhancements in the order of $5\dots6\cdot10^4$ were found. Looking at Eq.~\ref{eq:intermediate_intensities}, this implies an antenna efficiency of approximately $50\%$ in-between the nanospheres for the luminescence transition. This could be confirmed by simulations shown in Fig.~\ref{fig:antenna_efficiency}.

It can be stated that under the given assumptions, the dynamics
of the quantum system had to be taken into account to understand the
luminescence enhancement of the system driven by a quadrupolar excitation.
Only in the weak excitation limit, the luminescence enhancement simplifies to $\dot{n}_{1}^{\mathrm{rad},\mathrm{na}}/\dot{n}_{1}^{\mathrm{rad},\mathrm{fs}}
=\eta_{xz}^{\mathrm{loc}}\cdot\gamma_{10}^{\mathrm{rad,na}}/\gamma_{10}^{\mathrm{na}}$. From this relation it can clearly be seen that the local luminescence enhancement is directly proportional to the local quadrupole enhancement $\eta_{xz}^{\mathrm{loc}}$. For a sufficiently strong excitation field, however, the entire process saturates and the overall enhancement decreases.

\section{Conclusion}

In conclusion, it was shown that optical nanoantennas can effectively enhance higher order multipole transitions which are typically considered \textit{forbidden} in free space. This can be achieved by enhancing higher order multipole fields near the antenna. A quadrupole transition as the dominant excitation channel in a three-level system was considered. It was demonstrated how the enhancement of this transition can significantly intensify subsequent emission processes with respect to altered emission characteristics.
Since the effects under consideration depend on geometrical parameters, the properties of the optical nanoantenna can be tailored and hence allow for direct implementations in spectroscopic schemes. It must, furthermore, be emphasized that the example of the isolated dimer is just a specific case of what is possible with advanced optical nanoantennas. It can be anticipated that this work will give impetus for further research on such hybrid systems as well as plasmonic engineering.

\section*{Acknowledgements}

Financial support by the German Federal Ministry of Education and Research (PhoNa), by the Thuringian State Government (MeMa) and the German Science Foundation (SPP 1391 Ultrafast Nano-optics) is acknowledged. R.F. would like to thank B. Wellegehausen and T. Kienzler for helpful discussions and the anonymous referees for insightful comments.

\section*{Appendix}
\appendix

\section{Dipole vs. Quadrupole Enhancement\label{sub:Dipole_Enhancement}}

\begin{figure}
\begin{centering}
\includegraphics[width=40mm]{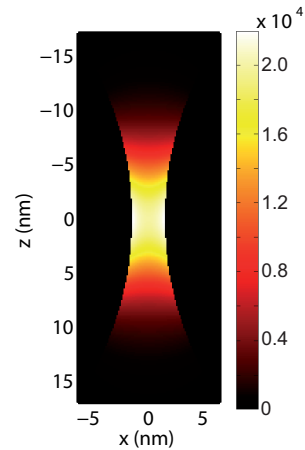}
\par\end{centering}
\caption{\label{fig:Dipole enhancement}
Local dipole enhancement factor $\eta_{x}^{\mathrm{loc,dip}}\left(\mathbf{r}_{0}\right)=\left|d_{x}^{\mathrm{na}}\left(\mathbf{r}_{0}\right)/d_{x}^{\mathrm{fs}}\left(\mathbf{r}_{0}\right)\right|^{2}$
for the main dipole contribution $d_{x}$. The configuration is the same dimer nanoantenna as used before with the same excitation wavelength $\lambda=437$~nm. The dipole enhancement is approximately two orders of magnitude smaller than the quadrupolar one.}
\end{figure}

In the considered model system it was assumed that the energies of each transition
are separated such that it is unlikely that different transitions
are competing for a certain excitation. So, only the excitation
rate with respect to the quantum system in free space was compared.
One may also ask the question what happens if two transitions,
namely a dipole and a quadrupole one, are energetically very close.
Can one then make the latter transition rate comparable or even stronger
than the dipole transition? This should
be possible when a certain eigenmode of a structure can be tailored
to have a strong quadrupole and only a negligible dipole field. For
the dimer antenna under consideration this is not the case and was not intended.
Here, also the electric dipole field gets enhanced. The result of a corresponding
calculation is outlined in Fig.~\ref{fig:Dipole enhancement}.
There one can see that the electric dipole field gets enhanced approximately
four orders of magnitude. This is still a few orders of magnitude
smaller than the quadrupolar one. However, the difference for this particular
antenna might not suffice to enhance a quadrupole transition to be stronger than
a dipolar one.

In general quantum systems, transitions might on the other hand be driven by a superposition of
several multipolar fields following Eq.~\ref{eq:Fermi}, i.e. for Rydberg atoms. An enhanced quadrupolar field component then might allow an additional energy supply channel to such a system. However, as in the discussed three-level system, an enhancement of higher order field componenents must be understood in terms of the whole dynamics of the quantum system. The approach might be most beneficial if this additional energy supply can account for an existing bottleneck for processes one wishes to enhance.

\section{Minimal Coupling}
In the main manuscript, the derivation of the minimal coupling potential $V_I=-\frac{e}{m}\mathbf{A}\cdot\mathbf{p}$ was not presented, as well as its decomposition into an electric and magnetic part given by Eq.~\ref{eq:el_magn_coupling}. Although, $V_I$ can be found in the literature and the derivation of its decomposition is straight forward, it will be outlined in the following two subsections.

\subsection{The Interaction Potential\label{sub:Minimal_Coupling_Interaction_Potential}}

The action of an electromagnetic field on a given system within the
framework of nonrelativistic quantum mechanics is realized due to a replacement in the unperturbed \noun{Hamiltonian} in the following way:
\begin{eqnarray*}
H_0\left(\mathbf{p},\mathbf{x}\right) & \rightarrow & H\left(\mathbf{p}-e\mathbf{A}\left(t,\mathbf{x}\right),\mathbf{x}\right)+eU\left(t,\mathbf{x}\right)\\
& \equiv & H_0 \left(\mathbf{p},\mathbf{x}\right) + V_{I} \left(\mathbf{p},\mathbf{x},t\right)
\end{eqnarray*}
which is called minimal coupling\cite{Hertel2007,Doughty1990}. Throughout the analysis, a hydrogen-like \noun{Hamiltonian} of the form
\begin{equation}
H_0\left(\mathbf{p},\mathbf{x}\right) = \frac{1}{2m}\mathbf{p}^{2}-\frac{1}{4\pi\epsilon_{0}}\frac{Z\,e^{2}}{r}
\end{equation} will be assumed.

The relation of the fields to the electrodynamic potentials is given by
\begin{eqnarray}
\mathbf{B}  =  \mathrm{curl}\mathbf{A} & \ \mathrm{and} \ &
\mathbf{E}  =  -\partial_{t}\mathbf{A}-\mathrm{grad}U\ .\label{eq:VectorPotential_Eqs}
\end{eqnarray}

Choosing the \noun{Coulomb} gauge in which
\begin{eqnarray}
\mathrm{div}\mathbf{A}  =  0 &\ \mathrm{and}\ & U \equiv 0\label{eq:Coulomb_gauge}
\end{eqnarray}
holds if $\rho=0$ and $\mathbf{j}=0$, one finds for the given \noun{Hamiltonian}
\begin{eqnarray}
H\left(\mathbf{p},\mathbf{x},t\right) & = & \frac{1}{2m}\left(\mathbf{p}-e\mathbf{A}\left(\mathbf{x},t\right)\right)^{2} - \frac{1}{4\pi\epsilon_{0}}\frac{Z\,e^{2}}{r} + eU\left(\mathbf{x},t\right)\nonumber
\\
& = & \frac{1}{2m}\left[ \mathbf{p}^{2} - e\mathbf{p}\cdot\mathbf{A} - e\mathbf{A}\cdot\mathbf{p} + e^{2}\mathbf{A}^{2} \right] -\frac{1}{4\pi\epsilon_{0}}\frac{Z\,e^{2}}{r}\nonumber
\\
& \approx & \frac{1}{2m}\mathbf{p}^{2}-\frac{1}{4\pi\epsilon_{0}}\frac{Z\,e^{2}}{r}-\frac{e}{m}\mathbf{A}\cdot\mathbf{p}\nonumber \\
& \equiv & H_{0}+V_I\ .\label{eq:Decomposition_Hamiltonian_Coulomb}
 \end{eqnarray}
where it was used that in the given gauge $U=0$ can always be achieved and $\mathbf{p}\cdot\mathbf{A}=\mathbf{A}\cdot\mathbf{p}$. Furthermore, the term in $\mathbf{A}^{2}$ has been neglected. This part corresponds to the \noun{Ponderomotive} force.

\subsection{Electric and Magnetic Coupling - a Decomposition\label{sub:Decomposition}}
Now that the interaction potential $V_I = -\frac{e}{m}\mathbf{A}\cdot\mathbf{p}$ is known, it is desirable to further split it into an electric and magnetic part. This approach eases later interpretation such as the attribution of transitions to electric and magnetic multipoles.

First of all, one finds
\begin{eqnarray*}
\mathbf{A}\cdot\mathbf{p}&=&\frac{\mathrm{i}m}{\hbar}\mathbf{A}\cdot\left[H_{0},\mathbf{x}\right]\\&=&\frac{\mathrm{i}m}{\hbar}\left\{ \mathbf{A}\cdot H_{0}\mathbf{x}-\mathbf{A}\cdot\mathbf{x}H_{0}+H_{0}\mathbf{A}\cdot\mathbf{x}-H_{0}\mathbf{A}\cdot\mathbf{x}\right\} \\&=&\frac{\mathrm{i}m}{\hbar}\left\{ \left[H_{0},\mathbf{A}\cdot\mathbf{x}\right]+\left[\mathbf{A},H_{0}\right]\mathbf{x}\right\} \ .
\end{eqnarray*}

Now, one can state for the first term
\begin{eqnarray*}
\left\langle m\right|\left[H_{0},\mathbf{A}\cdot\mathbf{x}\right]\left|n\right\rangle  & = & \hbar\omega_{mn}\left\langle m\right|\mathbf{A}\cdot\mathbf{x}\left|n\right\rangle \ .\end{eqnarray*}
Because of $\mathbf{E}=-\partial_{t}\mathbf{A}$, this term can be interpreted in a time-harmonic dependency of the fields as the electric contribution. Then, $\mathbf{E}=\mathrm{i}\omega\mathbf{A}$ and the electric coupling may be introduced as
\begin{eqnarray}
V_{e} & = & -\frac{e}{m}\frac{\mathrm{i}m}{\hbar}\left[H_{0},\mathbf{A}\cdot\mathbf{x}\right]\nonumber\\
 & = & -\frac{e}{\hbar\omega}\left[H_{0},\mathbf{E}\cdot\mathbf{x}\right]\ .\label{eq:Electric_Coupling}
\end{eqnarray}

On the other hand,
\begin{eqnarray*}
\left[\mathbf{A},H_{0}\right] & = & \frac{1}{2m}\left\{ \mathbf{A}\mathbf{p}^{2}-\mathbf{p}^{2}\mathbf{A}\right\} \\
 & = & -\frac{\hbar^{2}}{2m}\Delta\mathbf{A} = \frac{\hbar^{2}}{2m}\mathrm{curl}\mathbf{B}
 \end{eqnarray*}
where the curl is acting only on $\mathbf{B}$. It is further
 \begin{eqnarray*}
\left[\mathbf{A},H_{0}\right]\mathbf{x} & = & \frac{\hbar^{2}}{2m}\mathrm{curl}\mathbf{B}\cdot\mathbf{x}
  =  \frac{\hbar^{2}}{2m}\left(\mathbf{x}\times\nabla\right)\cdot\mathbf{B}\\
 & = & \mathrm{i}\frac{\mu_{B}}{e}\mathbf{L}\cdot\mathbf{B}
 \end{eqnarray*}
where $\mu_{B}=\frac{e\hbar}{2m}$, $\mathbf{p}=-\mathrm{i}\hbar\nabla$
and $\mathbf{L}=\mathbf{x}\times\mathbf{p}$ was used. So, it is natural
to define
\begin{eqnarray}
V_{m} & = & -\frac{e}{m}\cdot\frac{\mathrm{i}m}{\hbar}\cdot\mathrm{i}\frac{\mu_{B}}{e}\mathbf{L}\cdot\mathbf{B}
 = \frac{\mu_{B}}{\hbar}\mathbf{L}\cdot\mathbf{B}\label{eq:Magnetic_Coupling}
\end{eqnarray}
as the magnetic coupling.

\section{On the Choice of Energy Ranges and Decay Rates\label{sec:choices}}

The numbers chosen for the calculations were motivated by order-of-magnitude estimations for quantum
systems, namely hydrogen-like atoms and dye molecules. In this section, the choice of transition energies and used rates will be related to existing quantum systems. The suggested systems may be used in experiments.

\subsection*{Energy Ranges}

A quadrupolar transition can be seen as the sum over all possible
consecutive dipolar transitions. E.g. for the quadrupolar potential
$V\left(\mathbf{x}\right) \propto xy$ one finds
\begin{eqnarray*}
\Gamma_{ij}	&\propto&	\left|\left\langle i\right|xy\left|j\right\rangle \right|^{2}
	        =	\left|\sum_{k}\left\langle i\right|x\left|k\right\rangle \left\langle k\right|y\left|j\right\rangle \right|^{2}\ .
\end{eqnarray*}
Hydrogen-like atomic systems shall be considered now. There, such transitions may be realized by the transition from an s to a d orbital using a p orbital as main intermediate step. These three levels then form an effective three level system as the studied one.

The chosen energies correspond to such systems:
Potassium has a quadrupolar transition $\lambda_{4s \rightarrow 3d} \approx 446\,$nm
with intermediate step $\lambda_{4s \rightarrow 4p} \approx 770\,$nm followed by $\lambda_{4p \rightarrow 3d} \approx 1.18\,${\textmu}m.
The next alkali atom is Rubidium with $\lambda_{5s\rightarrow 4d} \approx 516\,$nm,
$\lambda_{5s\rightarrow 5p} \approx 780\,$nm and $\lambda_{5p\rightarrow 4d} \approx 1.4\,${\textmu}m. For Cesium, one finds $\lambda_{6s\rightarrow 5d} \approx 685\,$nm,
$\lambda_{6s\rightarrow 6p} \approx 894\,$nm and $\lambda_{6p\rightarrow 5d} \approx 2.9\,${\textmu}m.

However, Lithium and Sodium have higher quadrupolar transition energies with wavelengths around $300\,$nm. Thus for such atoms an experimental realization may not be feasible - gold and silver lose their metallic character for such high photon energies.

\subsection*{Decay Rates}

At a certain saturation intensity $I_{s}$, dipolar excitation rates become comparable to spontaneous
emission rates\cite{Metcalf1999,Novotny2006}. For a
two-level-system, $I_{s}$ is given by the intensity at which both
states are equally likely.

Hydrogen-like atoms exhibit an $I_{s}$ in the order of tens of $\text{W/m}^{2}$ at optical frequencies.
For instance, Rubidium has an $I_s = 16.4\,\text{W/m}^2$ characterizing the 5s-5p transition. On the other hand, $I_s$ may be more than $10^{7}\,\text{W/m}^{2}$ for dye molecules\cite{Novotny2006}.

At intensities above $I_{s}$, the excitation rates for dipolar excitations
become faster than their spontaneous counterparts. Quadrupolar
excitation rates are for optical frequencies in the order of six to
seven orders of magnitude smaller than dipolar ones. Thus, at intensities
$I_{0}\approx10^{3\dots4}I_{s}$, one might expect a quadrupole excitation
rate to be in the order of $\Gamma_{02}^{\mathrm{fs}}\left(I_{0}\right)\approx10^{-3}\gamma_{10}^{\mathrm{fs}}$
as was chosen in the calculations.

The normalized intensity range in Fig.~\ref{fig:Gamma_compare}~c), $I/I_{0}\approx10^{-4}\dots10^{4}$ corresponds to spectroscopic measurements.
Intensities employed in pulsed systems are in the order of $10^{12}\dots10^{14}\,\text{W/m}^{2}$
and for continuously operating systems with plasmonic structures $10^{6}\dots10^{8}\,\text{W/m}^{2}$.
Hence, for dye molecules pulsed measurements are at an intensity of
$I_{\mathrm{pulse}}\approx10^{6}I_{s}\approx10^{2\dots3}I_{0}$ and
continuous wave measurements are at $I_{\mathrm{cw}}\approx I_{s}\approx10^{-3\dots-4}I_{0}$.
As it was shown in Fig.~\ref{fig:Gamma_compare}~c), in-between these intensity ranges different limiting cases take place as discussed in section~\ref{sub:Excitation_Luminescence_Limiting}.

Furthermore, the fast nonradiative relaxation rate from state $\left|2\right\rangle $
to $\left|1\right\rangle $, $\gamma_{21}^{\mathrm{fs}}$ was assumed
to be the fastest process in free space. It was assumed to be independent of the environment,
$\gamma_{21}^{\mathrm{fs}}\equiv\gamma_{21}^{\mathrm{na}}=10^{2}\gamma_{10}^{\mathrm{fs}}$.
This assumption corresponds to fast thermal relaxations of excited atomic systems.


%

\end{document}